\documentclass[preprint,aps,superscriptaddress,nofootinbib,a4paper]{revtex4-1}
\usepackage[latin9]{inputenc}
\usepackage{verbatim}
\usepackage{amsmath}
\usepackage{amssymb}
\usepackage{graphicx}

\makeatletter

 \@ifundefined{textcolor}{}
 {%
   \definecolor{BLACK}{gray}{0}
   \definecolor{WHITE}{gray}{1}
   \definecolor{RED}{rgb}{1,0,0}
   \definecolor{GREEN}{rgb}{0,1,0}
   \definecolor{BLUE}{rgb}{0,0,1}
   \definecolor{CYAN}{cmyk}{1,0,0,0}
   \definecolor{MAGENTA}{cmyk}{0,1,0,0}
   \definecolor{YELLOW}{cmyk}{0,0,1,0}
 }

\usepackage{color}
\usepackage{braket}
\usepackage{mathrsfs}

\makeatother

\begin{document}

\ 

\

\preprint{NCTS-ECP/1601, MIT-CTP/4843}

\title{Low-energy electronic recoil in xenon detectors by solar neutrinos}

\author{Jiunn-Wei Chen\footnote{E-mail: jwc@phys.ntu.edu.tw}}

\affiliation{Department of Physics and Center for Theoretical Sciences, National Taiwan University, Taipei 10617, Taiwan}

\affiliation{Leung Center for Cosmology and Particle Astrophysics, National Taiwan
University, Taipei 10617, Taiwan}

\affiliation{Center for Theoretical Physics, Massachusetts Institute of Technology,
Cambridge, MA 02139, USA}

\author{Hsin-Chang Chi\footnote{E-mail: hsinchang@mail.ndhu.edu.tw}}

\affiliation{Department of Physics, National Dong Hwa University, Shoufeng, Hualien
97401, Taiwan}

\author{C.-P. Liu\footnote{E-mail: cpliu@mail.ndhu.edu.tw}}

\affiliation{Department of Physics, National Dong Hwa University, Shoufeng, Hualien
97401, Taiwan}

\author{Chih-Pan Wu\footnote{E-mail: d01222003@ntu.edu.tw}}

\affiliation{Department of Physics and Center for Theoretical Sciences, National
Taiwan University, Taipei 10617, Taiwan}
\begin{abstract}
Low-energy electronic recoil caused by solar neutrinos in multi-ton
xenon detectors is an important subject not only because it is a source
of the irreducible background for direct searches of weakly-interacting
massive particles (WIMPs), but also because it provides a viable way to
measure the solar $pp$ and $^{7}\textrm{Be}$ neutrinos at the precision
level of current standard solar model predictions. In this work we
perform \textit{ab initio} many-body calculations for the structure,
photoionization, and neutrino-ionization of xenon. It is found that
the atomic binding effect yields a sizable suppression to the neutrino-electron
scattering cross section at low recoil energies. Compared with the
previous calculation based on the free electron picture, our calculated
event rate of electronic recoil in the same detector configuration
is reduced by about $25\%$. We present in this paper the electronic
recoil rate spectrum in the energy window of 100 eV - %\textendash{}
30 keV with the standard per ton per year normalization for xenon
detectors, and discuss its implication for low energy solar neutrino
detection (as the signal) and WIMP search (as a source of background). 
\end{abstract}

\maketitle

\section{Introduction~\label{sec:intro}}

Direct searches of weakly-interacting massive particles (WIMPs), one
of the favored dark matter (DM) candidates, have been 
actively pursued in experimental nuclear and particle physics.
Although no concrete evidence of WIMPs has been obtained so far, a large portion of the parameter space 
(in terms of WIMP mass and their cross section to normal matter) has been ruled out. 
For example, recent results by the PandaX-II~\cite{Tan:2016zwf}
and LUX~\cite{Akerib:2016vxi} experiments, both employing xenon
detectors, set their best upper limits on the spin-independent WIMP-nucleon
cross section: $2.5\times10^{-46}\,\textrm{cm}^{2}$ for a $40\,\textrm{GeV}/c^{2}$
WIMP and $2.2\times10^{-46}\,\textrm{cm}^{2}$ for a $50\,\textrm{GeV}/c^{2}$
WIMP, respectively. 

Using Xenon as a detector has several advantages. It is relatively cheap to obtain, easy to scale up, and having enhanced cross
sections when scattered coherently. Therefore there are  several next-generation WIMP search proposals ---
XENON1T~\cite{Aprile:2015uzo}, LZ~\cite{Akerib:2015cja}, and DARWIN~\cite{Aalbers:2016jon}---
all use xenon as detectors. These are multi-ton scale detectors aiming at improving
the current sensitivity in WIMP-nucleon cross section by one order of magnitude with a
ton-year exposure (a modest goal) to three orders of magnitude with 200 ton-year exposure (an ambitious goal). 

To reach high sensitivity in those experiments, proper background removal is crucial. Direct WIMP searches
use nuclear recoil as a signal of WIMP-nucleus collision. However, nuclear recoil due to coherent neutrino-nucleus scattering could fake the signal. This kind of background is hard to shield and forms an irreducible
background called ``neutrino floor" which limits the ultimate sensitivity the experiments can achieve~\cite{Billard:2013qya}.

Neutrino electron scattering is another type of neutrino background which is in principle reducible, but in practice hard to remove completely in experiments. The DARWIN detector, for example, has only a small chance ($\sim 0.02 \%$) to mis-identify an electronic recoil in this process as a nuclear recoil signal. However, the large flux from $pp$ (end-point energy at 420 keV) and $^{7}\textrm{Be}$ (two discrete
energies at 862 and 384 keV) solar neutrinos makes electronic recoils the limiting background to measure the cross section on a nucleon lower than $4\times10^{-49}\,\textrm{cm}^{2}$ for WIMP mass of $40\,\textrm{GeV}/c^{2}$.

% because the WIMP detector does not have $100\%$ discriminating power between electronic and nuclear recoils. , electronic recoil has to be considered as a part of this irreducible background. Meanwhile, this is the main reaction channel that low energy solar neutrinos, i.e., the ones from $pp$ fusion (end-point energy at 420 keV) and $^{7}\textrm{Be}$ (two discrete energies at 862 and 384 keV), can be registered in a xenon detector with an energy threshold at about a few-keV level. In spite of a big cross section due to coherent scattering, the nuclear recoil, e.g., $\sim$16 eV maximum for a 1-MeV incident neutrino, is too small to be detected. Therefore, the low-energy recoil spectrum induced by solar neutrinos in xenon detectors not only is important for the background estimate in direct WIMP detection, more significantly, it can be used for precise $pp$ and $^{7}\textrm{Be}$ neutrino detection and checking the standard solar model predictions. (Note that the Borexino experiment observed only part of the $pp$~\cite{Bellini:2014uqa} and $^{7}\textrm{Be}$~\cite{Bellini:2011rx} neutrinos because of a high detection threshold at 50 keV, so its precisions in flux measurements are limited.) 

A very interesting observation made in Ref.~\cite{Baudis:2013qla} is that this very phenomenon of  neutrino electron scattering that limits the WIMP detection can be turned into an opportunity to measure low energy solar neutrino flux to high precision.  
% By assuming all electrons in a xenon atom are free particles while being scattered by solar neutrinos, Ref.~\cite{Baudis:2013qla} studied several neutrino physics topics in multi-ton liquid xenon detectors.
It was found that an integrated $pp$ neutrino rate of 5900 events
in the recoil energy window of 2-30 keV can be reached by a 70 ton-year
exposure, which provides the required statistics for a $1\%$-level
measurement in the $pp$ neutrino flux~\cite{Baudis:2013qla}. (Note that the Borexino experiment observed only part of the $pp$~\cite{Bellini:2014uqa} and $^{7}\textrm{Be}$~\cite{Bellini:2011rx} neutrinos because of a high detection threshold at 50 keV, so its precisions in flux measurements are limited.) 

However, as demonstrated in our previous work on germanium detectors~\cite{Chen:2013lba,Chen:2014dsa,Chen:2014ypv},
low-energy electronic recoil around a few keV starts to deviate from
the simple free electron approximation and the improved version by
including the stepping of atomic shells. In xenon detectors, one expects
a similar and even bigger effect from atomic binding. To address
this important issue, we adopt an \textit{ab initio} many-body method:
the relativistic random phase approximation (RRPA)~\cite{Johnson:1979wr,Johnson:1979ch,Huang:1981jc}.
We first benchmark our calculations with Xenon structure and photo absorption data, then give a reliable prediction for the low-energy electronic recoil spectrum induced by solar neutrinos. 

The organization of this paper is as follows. In Sec.~\ref{sec:formalism},
we gather the essential formalism for neutrino-ionization of an atom
and its corresponding electronic recoil spectrum. In Sec.~\ref{sec:benchmark},
we justify our many-body approach to the xenon atom by showing benchmark
results on atomic structure and photoabsorption calculations. Our
main results for the electronic recoil spectrum induced by solar neutrinos
in multi-ton-scale xenon detectors are presented and discussed in
Sec.~\ref{sec:r=000026d}. Then we summarize in Sec.~\ref{sec:sum.}.

\section{General Formalism~\label{sec:formalism}}

To calculate the differential cross section for neutrino-ionization
of xenon atoms, caused by the weak neutrino-electron interactions,
we follow the general formalism described in Ref.~\cite{Chen:2013lba,Chen:2014dsa,Chen:2014ypv}.
In the massless limit of neutrinos $m_{\nu}\rightarrow0$, the differential
cross section with respect to the energy deposition, denoted as $T$,
by the incoming neutrino of flavor $i$, is 
\begin{eqnarray}
\dfrac{d\sigma^{(i)}}{dT} & = & \dfrac{G_{F}^{2}}{\pi}(E_{\nu}-T)^{2}\int d\cos\theta\,\cos^{2}\dfrac{\theta}{2}\Big\{ R_{00}^{(i)}-\dfrac{T}{|\vec{q}|}R_{03+30}^{(i)}+\dfrac{T^{2}}{|\vec{q}|^{2}}R_{33}^{(i)}\nonumber \\
 & + & \left(\tan^{2}\dfrac{\theta}{2}-\dfrac{q^{2}}{2|\vec{q}|^{2}}\right)R_{11+22}^{(i)}+\tan\dfrac{\theta}{2}\sqrt{\tan^{2}\dfrac{\theta}{2}-\dfrac{q^{2}}{|\vec{q}|^{2}}}R_{12+21}^{(i)}\Big\}\label{eq:dS/dT_weak}
\end{eqnarray}
in the laboratory frame, where $G_{F}$ is the Fermi constant; $E_{\nu}$
the incident neutrino energy; $\theta$ the neutrino scattering angle;
and $q$ ($\vec{q}$) the 4- (3-) momentum transfer of the neutrino,
respectively. The atomic weak response functions 
\begin{eqnarray}
R_{\mu\nu}^{(i)} & = & \frac{1}{2J_{i}+1}\sum_{M_{J_{i}}}\sum_{f}\braket{\Psi_{f}|c_{V}^{(i)}\hat{\mathcal{J}}_{\mu}-c_{A}^{(i)}\hat{\mathcal{J}}_{5\mu}|\Psi_{i}}\braket{\Psi_{f}|c_{V}^{(i)}\hat{\mathcal{J}}_{\nu}-c_{A}^{(i)}\hat{\mathcal{J}}_{5\nu}|\Psi_{i}}^{*}\nonumber \\
 &  & \times\delta(T+E_{i}-E_{f})\,,\label{eq:RF_weak}
\end{eqnarray}
with Lorentz indices $\mu,\nu=0,1,2,3$ (the 3-axis is defined by
the direction of $\vec{q}$) involve a sum of the final atomic states
$\ket{\Psi_{f}}$ and a spin average of the initial atomic states
$\ket{\Psi_{i}}=\ket{J_{i},M_{J_{i}},\ldots}$, and the Dirac delta
function imposes energy conservation. The (axial-) vector current
operator for an electron filed $\hat{\psi}_{e}$ is represented in
momentum space 

\begin{equation}
\hat{\mathcal{J}}_{\mu(5)}\equiv\int d^{3}x\,e^{i\vec{q}\cdot\vec{x}}\hat{\bar{\psi}}_{e}(\vec{x})\gamma^{\mu}(\gamma_{5})\hat{\psi}_{e}(\vec{x})\,.\label{eq:J_atom}
\end{equation}
Depending on the flavor of the incident neutrino, the vector and axial-vector
coupling constants are 
\begin{align}
c_{V}^{(i)}=-\frac{1}{2}+2\sin^{2}\theta_{w}+\delta_{i,e}\,, & \qquad c_{A}^{(i)}=-\frac{1}{2}+\delta_{i,e}\,,\label{eq:coupling}
\end{align}
where $\theta_{w}$ is the Weinberg angle; and the difference between
$\nu_{e}$ and $\nu_{\mu,\tau}$ is because the former scattering
involves both the charged and neutral weak interactions, while the
latter is purely neutral. 

The differential electronic recoil spectrum induced by a neutrino
source is calculated by folding the above differential cross sections
$d\sigma^{(i)}(T,E_{\nu})/dT$ with the incident neutrino energy spectrum
$d\phi^{(i)}(E_{\nu})/dE_{\nu}$: 
\begin{equation}
\frac{dN_{e}(E_{e})}{dT}=N_{0}\times t\times\sum_{i=e,\mu,\tau}\int dE_{\nu}\frac{d\phi^{(i)}(E_{\nu})}{dE_{\nu}}\frac{d\sigma^{(i)}(E_{e},E_{\nu})}{dT}\,.
\end{equation}
We shall adopt the standard per-ton per-year per-keV normalization,
so the total number of atoms $N_{0}=6.02\times10^{29}/A$ ($A$ is
the atomic mass of the detector atom in atomic units), $t$ is one
year, and energy is measured in units of keV. 

\section{Benchmark Calculations of The Xenon Atom \label{sec:benchmark}}

\begin{table*}[t]
\caption{The single-particle energies of a Xe atom calculated by DHF (s.p.)
in this work versus the edge energies extracted from photoabsorption
data (edge) in Ref.~\cite{Henke:1993gd} (The one for the $K$-shell
is not available). All energies are in units of $\mathrm{eV}$. \label{tab:sp_energy}}

\begin{ruledtabular}
\begin{tabular}{cccccccccc}
 & $K(1s_{\frac{1}{2}})$ & $L_{I}(2s_{\frac{1}{2}})$ & $L_{II}(2p_{\frac{3}{2}})$ & $L_{III}(2p_{\frac{1}{2}})$ & $M_{I}(3s_{\frac{1}{2}})$ & $M_{II}(3p_{\frac{3}{2}})$ & $M_{III}(3p_{\frac{1}{2}})$ & $M_{IV}(3d_{\frac{5}{2}})$ & $M_{V}(3d_{\frac{3}{2}})$\tabularnewline
s.p. & 34759.3 & 5509.8 & 5161.5 & 4835.6 & 1170.5 & 1024.8 & 961.2 & 708.1 & 694.9\tabularnewline
edge & - & 5452.8 & 5103.7 & 4782.2 & 1148.7 & 1002.1 & 940.6 & 689.0 & 676.4\tabularnewline
\hline 
 & $N_{I}(4s_{\frac{1}{2}})$ & $N_{II}(4p_{\frac{3}{2}})$ & $N_{III}(4p_{\frac{1}{2}})$ & $N_{IV}(4d_{\frac{5}{2}})$ & $N_{V}(4d_{\frac{3}{2}})$ & $O_{I}(5s_{\frac{1}{2}})$ & $O_{II}(5p_{\frac{1}{2}})$ & $O_{III}(5p_{\frac{3}{2}})$ & \tabularnewline
s.p. & 229.4 & 175.6 & 162.8 & 73.8 & 71.7 & 27.5 & 13.4 & 12.0 & \tabularnewline
edge & 213.2 & 146.7 & 145.5 & 69.5 & 67.5 & 23.3 & 13.4 & 12.1 & \tabularnewline
\end{tabular}
\end{ruledtabular}

\end{table*}

\begin{figure}[h]
 \includegraphics[width=0.72\textwidth]{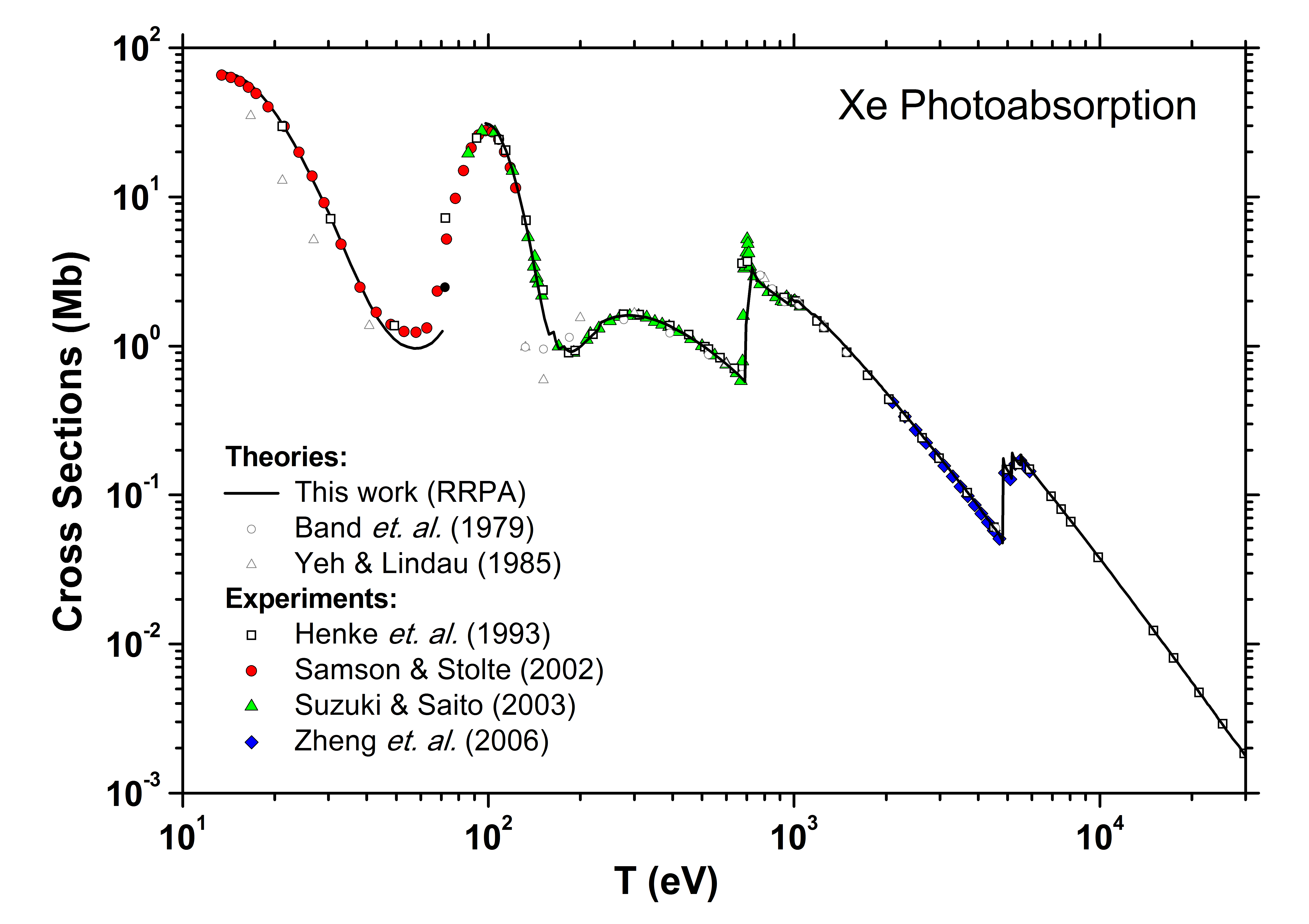}
\includegraphics[width=0.72\textwidth]{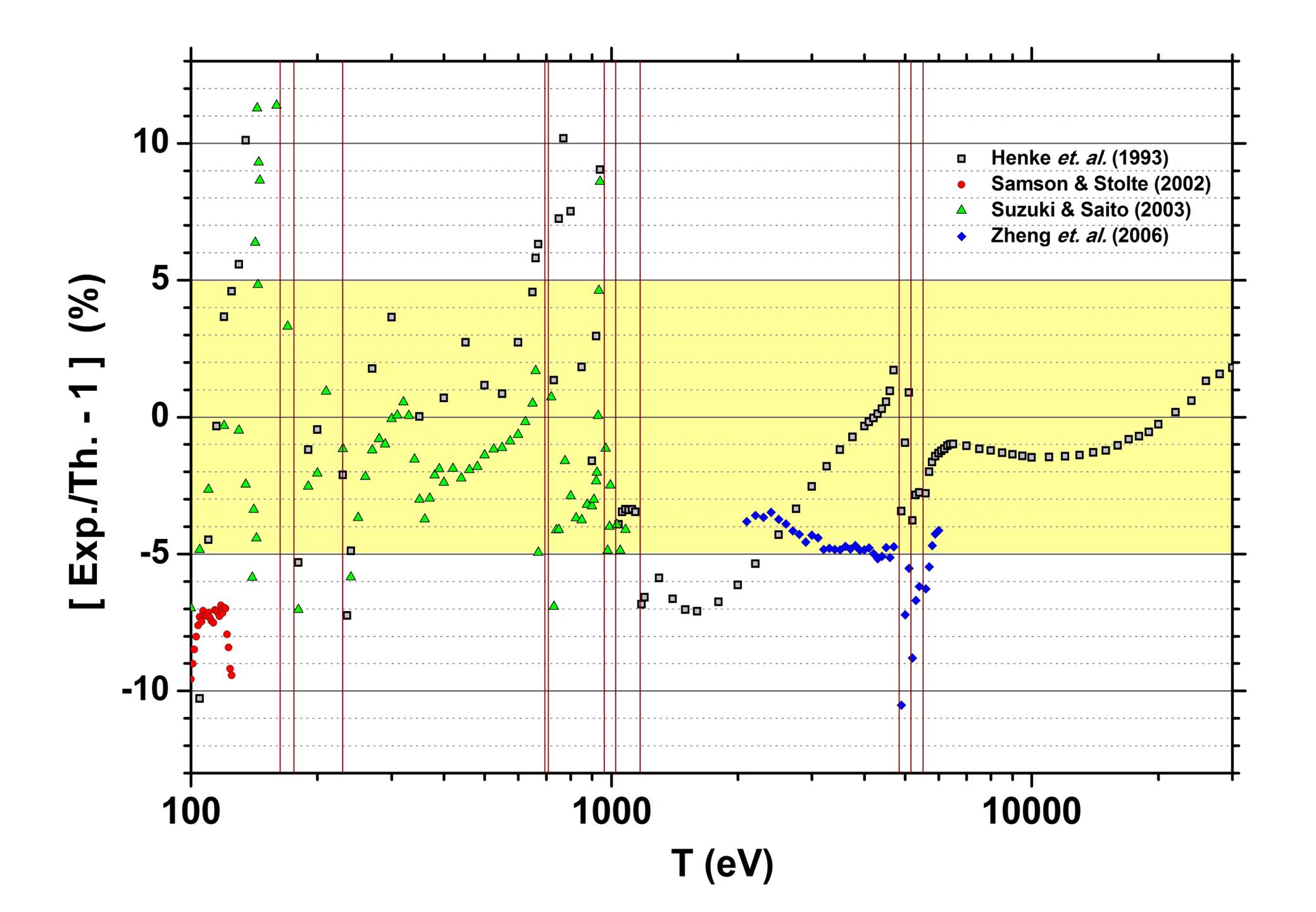}
\caption{(Top) Photoabsorption cross section of Xe. The black solid line shows
the results of our RRPA calculation. Points in empty squares and colors
are experimental data compiled from Refs.~ \cite{Henke:1993gd,Samson:2002xe,Suzuki:2003xe,Zheng:2006xe},
and points in empty circles and triangles are theory predictions~\cite{Band:1979xe,Yeh:1985xe}.
(Bottom) The relative difference between our RRPA calculation and
experimental data. \label{fig:photo_abs} }
\end{figure}

The atomic many-body wave functions for the initial and final states
are computed by an \textit{ab initio} method: the relativistic random
phase approximation (RRPA)~\cite{Johnson:1979wr,Johnson:1979ch,Huang:1981jc}.\footnote{In our previous many-body calculations for germanium atoms \cite{Chen:2014ypv},
we need one extra feature, the multiconfiguration ground state, to
handle open-shell atoms properly. The method was given the name: multiconfiguration
random phase approximation (MCRRPA) by its early pioneers \cite{Johnson:1979wr,Huang:1981wj}.} Because the atomic number of xenon, $Z=54$, is large, both the relativistic
effect and residual two-electron correlation are important. Therefore,
the RRPA provides the essential improvement over the Hartree-Fock
theory (a nonrelativistic mean-field theory) in obtaining good-quality
xenon wave functions of not only the ground state but also excited
states. 

Benchmark calculations for our atomic many-body calculations of xenon
were done in two steps. First we compare in Table~\ref{tab:sp_energy}
all the single-particle energies calculated by the Dirac-Hartree-Fock
(DHF) theory with the edge energies extracted from photoabsorption
data accumulated before 1990~\cite{Henke:1993gd}. Except for a few
intermediate shells including $N_{I}$, $N_{II}$, $N_{III}$, and
$O_{I}$, the general agreement is good. %\Blue{Effect?}
%As we will see later that these discrepancy will not have much effect on the solar neutrino electron scattering rate

Second, we calculate the total cross section of photoabsorption $\sigma_{\gamma}$,
which is dominated by photoionization for photon energy ranging between
12.1 eV (the ionization threshold) and 30 keV. Our result is shown as the black solid
line in the top panel of Fig.~\ref{fig:photo_abs} and is compared with experimental data compiled from Refs.~\cite{Henke:1993gd,Samson:2002xe,Suzuki:2003xe,Zheng:2006xe}.
The computation does not converge well between 70 -100 eV, so it is
left empty in this range except the black point just above 70 eV.
Our result visibly deviates from data between 40 -70 eV, but generally
agrees with data well across more than four orders of magnitude 
in cross section. Two previous theory predictions~\cite{Band:1979xe,Yeh:1985xe}
are also shown in this panel by empty circles and triangles, respectively.
In comparison, our calculation consistently works better in this broad
energy range considered.

More detailed comparison between different sets of data and our result
is shown in the the bottom panel of Fig.~\ref{fig:photo_abs}. The
data have at least $\sim2-5\%$ (1 sigma) errors which are not shown
here. In most region, the difference is less than $5\%$. Larger differences
could happen near ionization thresholds of atomic shells (indicated
by vertical lines), but they do not give significant contributions
when a broad range of spectrum is integrated. Therefore, we can assign
a conservative averaged theory error of 5\% to our calculation in
the energy range of $100\,\textrm{eV}\leq T\leq30\,\textrm{keV}$.
If one only considers the energy range between 2\textendash 30 keV,
the averaged error is further reduced to $2-3\%$. 

\section{Results and Discussion~\label{sec:r=000026d}}

Taking confidence on the xenon wave functions obtained by our many-body
approach, the neutrino-ionization process is computed as outlined
in Eqs.~(\ref{eq:dS/dT_weak}\textendash \ref{eq:coupling}) in the
energy range of $100\,\textrm{eV}\le T\le30\,\textrm{keV}$.

For solar neutrino flux, we consider two main sources: the proton-proton
fusion, $p+p\rightarrow d+e^{+}+\nu_{e}$ (the $pp$ neutrinos), and
the electron capture by $^{7}\textrm{Be}$, $^{7}\textrm{Be}+e^{-}\rightarrow{}^{7}\textrm{Li}+\nu_{e}$
(the $^{7}\textrm{Be}$ neutrinos). The former has a continuous spectrum
ended at $420\,\textrm{keV}$; the latter has two discrete spectral
lines: one at $862\,\textrm{keV}$ and the other at $384\,\textrm{keV}$
with branching ratios $89.6\%$ and $10.4\%$, respectively. Together
they amount to $98\%$ of the total solar neutrino flux. 

The fluxes of the $pp$ and $^{7}\textrm{Be}$ neutrinos 
\begin{equation}
\phi_{pp}=5.98\times10^{10}\,\textrm{cm}^{-2}\,\textrm{s}^{-1},\;\phi_{^{7}\textrm{Be}}=5.00\times10^{9}\,\textrm{cm}^{-2}\,\textrm{s}^{-1}\,,
\end{equation}
are taken from the recent Standard Solar Model (SSM) prediction in
\cite{Serenelli:2011py}, which incorporated updated nuclear reaction
rates in \cite{Adelberger:2010qa}.\footnote{The $pp$ and $^{7}\textrm{Be}$ neutrino fluxes adopted in \cite{Baudis:2013qla}
is from an earlier SSM prediction \cite{Bahcall:2004pz}; both are
somewhat smaller than the ones of \cite{Serenelli:2011py}.} The spectral shape of the $pp$ neutrinos is approximated by the
standard $\beta$-decay form 
\begin{equation}
\dfrac{d\phi_{pp}^{(e)}(E_{\nu})}{dE_{\nu}}=A(Q+m_{e}-E_{\nu})[(Q+m_{e}-E_{\nu})^{2}-m_{e}^{2}]^{\frac{1}{2}}E_{\nu}^{2}F\,,
\end{equation}
where the $Q$-value is $420\,\textrm{keV}$, the Fermi function $F\cong1$
, and the normalization factor $A=2.97\times10^{-36}\,\textrm{keV}^{-2}$
is fixed by the chosen $pp$ flux $\phi_{pp}^{(e)}$ above.\footnote{The $pp$ flux is peaked within 100-400 keV in neutrino energy. 
This  simple parametric form agrees within 1\% with the more
sophisticated one derived from an explicit solar model calculation
\cite{Bahcall:1997eg} for $E_{\nu}\gtrsim10\,\textrm{keV}$. }

The flavor content of solar neutrinos seen in terrestrial detectors
is modified by neutrino oscillation both in vacuum between the Earth
and the surface of the Sun, and in medium between the solar surface
and core by the Mikheyev-Smirnov-Wolfenstein (MSW) effect. Current
observations of low-energy solar neutrinos including $pp$ and $^{7}\textrm{Be}$
are consistent with the MSW\textendash large-mixing-angle (MSW-LMA)
solution: It predicts a vacuum-dominated oscillation pattern and the
survival probability of electron neutrinos can be approximated as~\cite{Bahcall:2004mz}
\begin{equation}
P_{ee}=\cos^{4}\theta_{13}\left(1-\frac{1}{2}\sin^{2}(2\theta_{12})\right)+\sin^{4}\theta_{13}\,.
\end{equation}
Using the most recent oscillation angles recommended by the Particle
Data Group: $\sin^{2}(2\theta_{12})=0.846\pm0.021$ and $\sin^{2}(2\theta_{13})=0.085\pm0.005$~\cite{Agashe:2014kda},
$P_{ee}=0.553$ with an error about $2\%$. The remaining part of
the $pp$ and $^{7}\textrm{Be}$ fluxes contains either $\nu_{\mu}$
or $\nu_{\tau}$, and scatters with the same differential cross section
formula $d\sigma^{(\mu)}/dT=d\sigma^{(\tau)}/dT$.

\begin{figure}[h]
\includegraphics[width=0.75\textwidth]{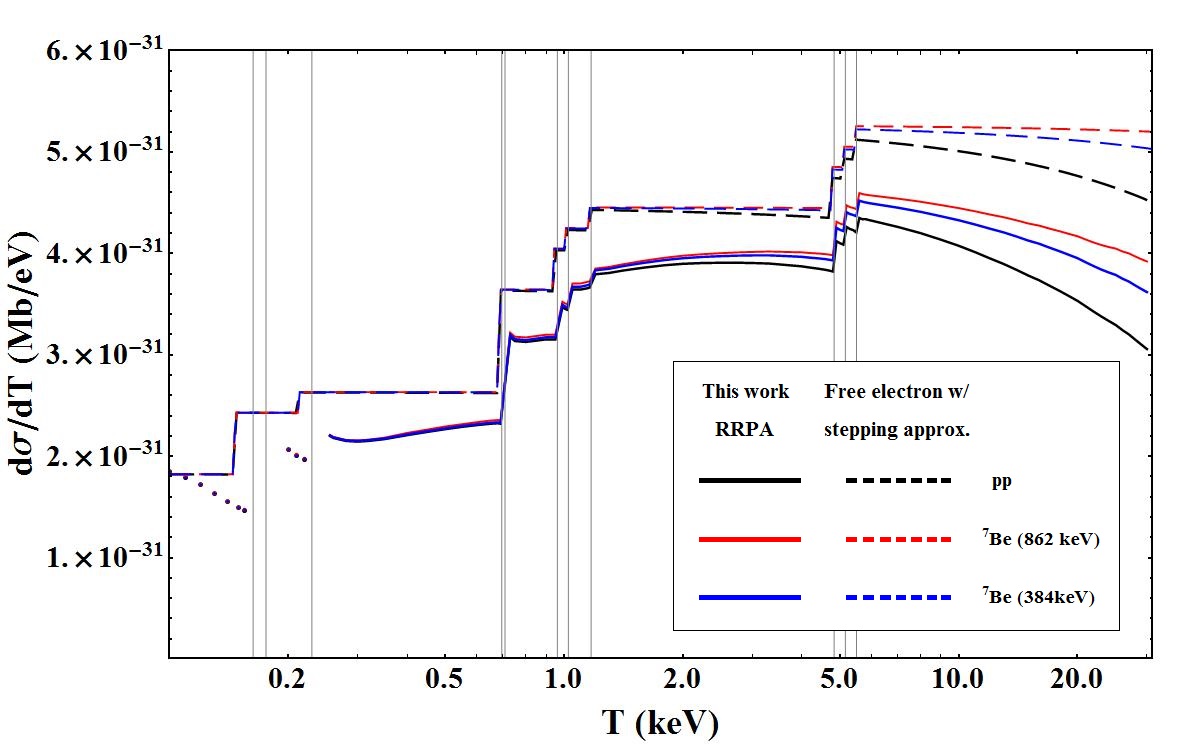}
\caption{Flux-averaged differential cross sections of xenon ionization by electron
neutrinos from $pp$, $^{7}\textrm{Be}(862\,\textrm{keV})$, and $^{7}\textrm{Be}(384\,\textrm{keV})$
sources in solid black, red, and blue lines, respectively. The dashed
lines are the predictions of the stepping approximation (see text
for more details). \label{fig:nu-Xe}}
\end{figure}

In Fig.~\ref{fig:nu-Xe}, we show our flux-averaged $\nu_{e}$-Xe
differential cross sections (in solid lines) for electron neutrinos
of $pp$, $^{7}\textrm{Be}(862\,\textrm{keV})$, and $^{7}\textrm{Be}(384\,\textrm{keV})$,
respectively. The general trends of $\left\langle d\sigma/dT\right\rangle $
show little dependence on neutrino sources, and its functional behavior
is largely controlled by the value of $T$ and the binding energies
of atomic shells (indicating by the vertical thin lines). The former
determines the number of electrons that can be ionized, and the latter
gives those sharp edges indicating large differential cross sections
whenever an atomic shell is just open. In combination, the largest
value of $\left\langle d\sigma/dT\right\rangle $ is reached at $T\sim5\,\textrm{keV}$,
which corresponds to the opening of $L$ shells of xenon. 

The dashed lines in this figures are the predictions of the stepping
approximation 
\begin{equation}
\frac{d\sigma^{(i)}}{dT}=\sum_{i=1}^{Z}\theta(T-B_{i})\frac{d\sigma_{0}^{(i)}}{dT}\,.
\end{equation}
This is done by weighting the scattering cross section of a neutrino
and a free electron, $d\sigma_{0}/dT$, with the number of electrons
that can be ionized by an energy deposition of $T$, where $\theta$
is the step function and $B_{i}$ is the binding energy of the $i$th
electron. It is clearly shown in these figures that the atomic binding
suppresses the cross sections, similar to our previous work on the
germanium atom.

Note that in this plot of $\left\langle d\sigma/dT\right\rangle $
we do not try to join the points at low recoil energies, $T\lesssim250\,\textrm{eV}$.
This reflects a numerical difficulty of obtaining stable solutions
near the low-energy edges. In general, one can expect a sharp increase
in $\left\langle d\sigma/dT\right\rangle $ when an atomic shell is
just open, and use the stepping approximation prediction to set a
upper bound. For more reliable predictions from detailed
many-body calculations, we shall leave them to future studies. 

\begin{figure}[h]
 \includegraphics[width=0.75\textwidth]{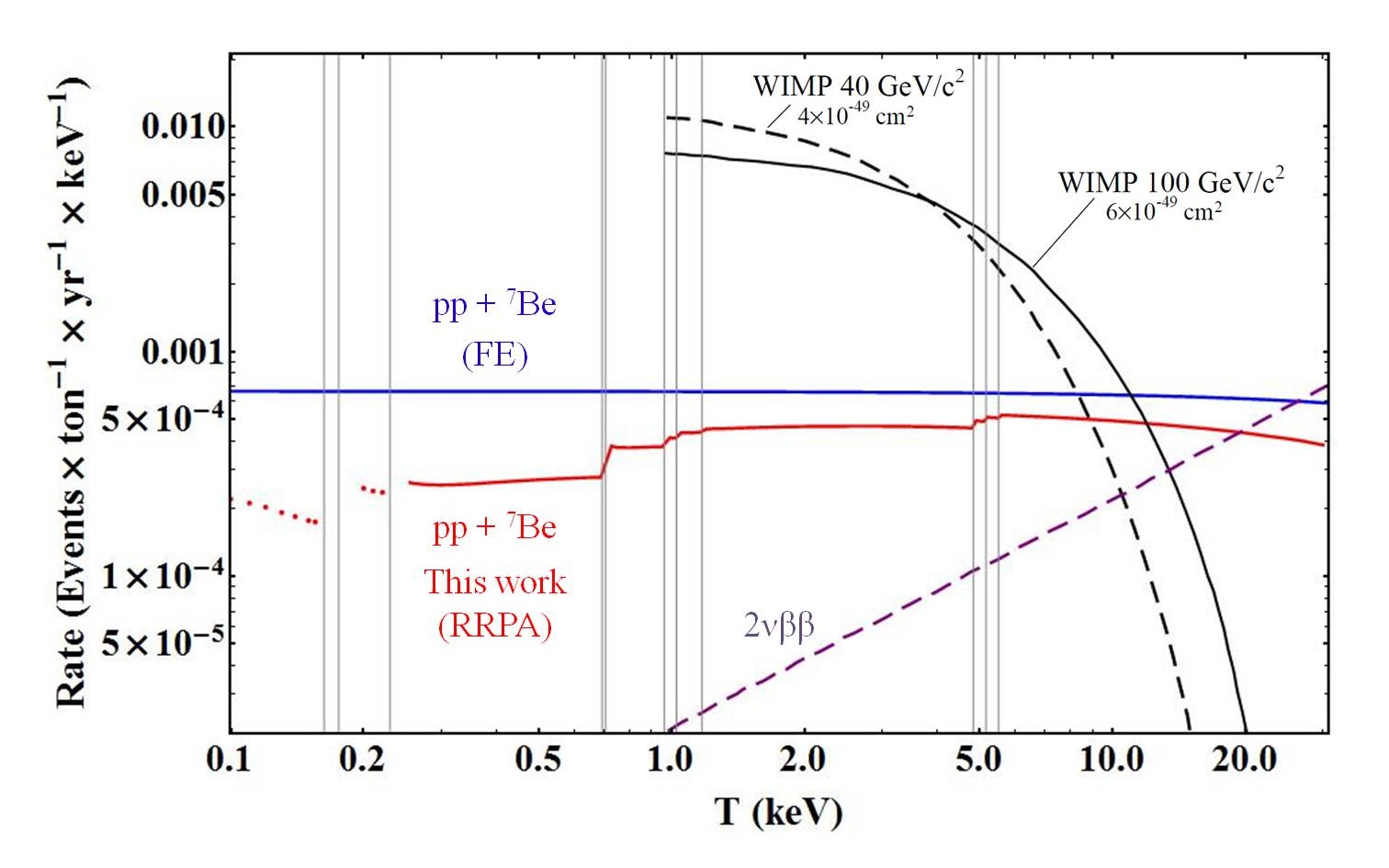}
\caption{Differential count rate of electronic recoil induced by solar neutrinos
in xenon detectors assuming a $99.98\%$ rejection. The red line is
our RRPA result and the blue line 
is the free electron approximation without stepping function as adopted in Ref.~\cite{Aalbers:2016jon}. Also shown
are another neutrino background from the two-neutrino double $\beta$ decay ($2\nu\beta\beta$) of $^{136}$Xe
and two cases of WIMP induced nuclear recoil (with DARWIN detector parameters) translated into electron
equivalent energy. The $2\nu\beta\beta$ and WIMP curves all taken from Ref.~\cite{Aalbers:2016jon}. \label{fig:dRdT}}
\end{figure}

Finally we present in Fig.~\ref{fig:dRdT} the differential count
rate of electronic recoil induced by solar neutrinos in a xenon detector
of one-ton mass and one-year exposure and assuming a $99.98\%$ event
rejection \cite{Aalbers:2016jon}.
%(Note that this level of rejection was achieved by XENON100~\cite{Aprile:2012nq} and used for discussions in Ref.~\cite{Baudis:2013qla}. 
(Other rejection levels of $99.6\%$~\cite{Akerib:2013tjd}
and $99.987\%$~\cite{Akimov:2011tj} are used for discussion in
Ref.~\cite{Schumann:2015cpa}). The energy range, $100\,\textrm{eV}$\textendash $30\,\textrm{keV}$
is fixed by the threshold of detecting electronic recoil at the low
end\footnote{We thank H. Nelson for pointing to us this low threshold of xenon
detectors in electronic recoil.} and the dominance of the $2\nu\beta\beta$ decay background from $^{136}$Xe 
%, $^{85}$Kr, and $^{222}$Rn
(the dotted
blue line) at the high end. Compared with Ref.~\cite{Aalbers:2016jon},
which used a naive free electron approximation without stepping (i.e.,
all 54 electrons of xenon contribute regardless of $T$), our calculation
show a consistent suppression due to atomic binding (the red vs. the
blue line). Also shown
are two cases of WIMP induced nuclear recoil (with DARWIN detector parameters) translated into electron
equivalent energy to show that solar neutrino background becomes the limiting factor in these scenarios. The $2\nu\beta\beta$ and WIMP curves all taken from Ref.~\cite{Aalbers:2016jon}.

In Table~\ref{tab:comparison}, we compare our predictions of solar neutrino induced
event rates, assuming a 1-ton-year exposure, with the ones of the free
electron (FE) approximation without stepping functions introduced, as
adopted in Refs.~\cite{Aalbers:2016jon,Baudis:2013qla}.\footnote{We note that our FE results differ slightly from Refs.~\cite{Aalbers:2016jon,Baudis:2013qla}
due to different mixing parameters, neutrino fluxes, etc. being used.
Also we would like to point out a typographical error in Eq.(2.5) of
Ref.~\cite{Baudis:2013qla}: the last term should be $g_L g_R \frac{m_e T}{E^2_{\nu}} $.}
%In Table~\ref{tab:comparison}, we compare our predictions of solar neutrino induced event rates with the free electron approximation (without stepping function) adopted in Ref.~\cite{Aalbers:2016jon} for 1-ton-year exposure. 
Using this as a channel for
low energy solar neutrino detection, the atomic binding effect reduces
the event rates in the energy window of 2-30 keV by $28\%$ and $24\%$
respectively for $pp$ and $^{7}\textrm{Be}$ neutrinos. This causes
a huge loss of statistics that is required for checking the standard
solar model prediction of low-energy solar neutrino fluxes at $1\%$ level. On the other hand, our theoretical
calculation is benchmarked with an error estimate at $2-3\%$ level,
which is just slightly larger than the desired goal, so the experimental
data can be interpreted without this theoretical uncertainty under
good control. As for this channel being a potential background in
WIMP searches, the suppression due to atomic binding turns out to
be a good news: It implies the sensitivity to WIMP-nucleon cross section
is increased by a similar factor, unless the rejection of electronic
recoil can reach an even higher level such that this background
becomes subdominant compared to the one from coherent neutrino-nucleus
scattering.

\begin{table}
\caption{Comparison of solar neutrino induced event rates predicted by our
many-body calculations (RRPA) and the free electron (FE) approximation
without stepping functions (as adopted by Refs.~\cite{Aalbers:2016jon,Baudis:2013qla}) for a xenon
detector with 1-ton-year exposure.
%Comparison of solar neutrino induced event rates predicted in this work (RRPA) and Ref.~\cite{Aalbers:2016jon} (free electron (FE) approximation without stepping function) for a xenon detector with 1-ton-year (1 t.y) exposure. 
\label{tab:comparison}}

\begin{tabular}{|lrrcrrc|}
\hline 
Physics channel & \multicolumn{3}{r}{Low-energy $\nu$ measurement} & \multicolumn{3}{c|}{Dark matter search}\tabularnewline
Energy range & \multicolumn{3}{c}{2\textendash 30 keV} & \multicolumn{3}{c|}{2\textendash 10 keV}\tabularnewline
Assumptions & \multicolumn{3}{c}{No ER/NR discrimination} & \multicolumn{3}{c|}{$99.98\%$ ER rejection, $30\%$ NR acceptance}\tabularnewline
\hline 
Model & FE & RRPA & Diff. & FE & RRPA & Diff.\tabularnewline
Solar $pp$ neutrinos & 80.7 & 57.8 & $-28\%$ & $4.8 \times 10^{-3}$ & $3.6 \times 10^{-3}$ & $-25\%$\tabularnewline
Solar $^{7}\textrm{Be}$ neutrinos & 7.1 & 5.4 & $-24\%$ & $4.1 \times 10^{-4}$ & $3.2 \times 10^{-4}$ & $-23\%$\tabularnewline
Total & 87.8 & 63.2 & $-28\%$ & $5.2 \times 10^{-3}$ & $3.9 \times 10^{-3}$ & $-24\%$\tabularnewline
\hline 
\end{tabular}
\end{table}

\section{Summary~\label{sec:sum.}}

The importance of low-energy electronic recoil induced by solar neutrinos
in multi-ton xenon detector is two-fold: On one hand, it is a
background that should be properly removed in searches of WIMP-nucleus
scattering. On the other hand, it provides a viable channel to detect
solar $pp$ and $^{7}\textrm{Be}$ in ``real time" \cite{Aalbers:2016jon} and has the potential
to measure their fluxes at the precision level of standard solar model
predictions. In this work, we have applied an \textit{ab initio} many-body
method: the relativistic random phase approximation (RRPA) to this
problem with good benchmarks from xenon structure and photoabsorption
data. We have also presented the electronic recoil rate spectrum
in the energy window of 100 eV \textendash{} 30 keV with the standard
per ton per year normalization for xenon detectors. Except small regions
near ionization thresholds of atomic shells, we estimate our theoretical
error less than $5\%$ in general, and the averaged error in the energy
window of 2\textendash 30 keV only $2-3\%$, which is just slightly
larger than the precision level ($\sim1\%$) of the current solar models.
We found that the atomic binding effect yields a sizable suppression
to the neutrino-electron scattering cross section at low recoil energies
in xenon detectors. Compared with the previous calculation based on
the free electron picture, our calculated event rate of electronic
recoil in the same detector configuration is reduced by about $30\%$.
This increases the demand of detector volume and/or exposure time
for precision measurements of solar neutrino fluxes. However, it also increases the
sensitivity to WIMP-nucleon scattering cross section.
\begin{acknowledgments}
We thank Dr. Harry Nelson (UCSB) for bringing this interesting problem
to our attention and useful discussions. We also thank Chih-Liang Wu for early involvement in this project.
This work is supported in
part by the Ministry of Science and Technology, Taiwan under Grants
Nos. 105-2112-M-002-017-MY3 and 105-2918-I-002-003 (J.-W. C. and C.-P.
W.), 104-2112-M-259-004-MY3 (C.-P. L.); the National Center for Theoretical
Sciences; the Center for Theoretical Sciences and Center of Advanced
Study in Theoretical Sciences of National Taiwan University (J.-W.
C. and C.-P. W.); and MIT MISTI program (J.-W. C. and C.-P. W.). 
\end{acknowledgments}

\bibliographystyle{apsrev4-1}
\bibliography{nuXe}

\begin{thebibliography}{31}%
\makeatletter
\providecommand \@ifxundefined [1]{%
 \@ifx{#1\undefined}
}%
\providecommand \@ifnum [1]{%
 \ifnum #1\expandafter \@firstoftwo
 \else \expandafter \@secondoftwo
 \fi
}%
\providecommand \@ifx [1]{%
 \ifx #1\expandafter \@firstoftwo
 \else \expandafter \@secondoftwo
 \fi
}%
\providecommand \natexlab [1]{#1}%
\providecommand \enquote  [1]{``#1''}%
\providecommand \bibnamefont  [1]{#1}%
\providecommand \bibfnamefont [1]{#1}%
\providecommand \citenamefont [1]{#1}%
\providecommand \href@noop [0]{\@secondoftwo}%
\providecommand \href [0]{\begingroup \@sanitize@url \@href}%
\providecommand \@href[1]{\@@startlink{#1}\@@href}%
\providecommand \@@href[1]{\endgroup#1\@@endlink}%
\providecommand \@sanitize@url [0]{\catcode `\\12\catcode `\$12\catcode
  `\&12\catcode `\#12\catcode `\^12\catcode `\_12\catcode `\%12\relax}%
\providecommand \@@startlink[1]{}%
\providecommand \@@endlink[0]{}%
\providecommand \url  [0]{\begingroup\@sanitize@url \@url }%
\providecommand \@url [1]{\endgroup\@href {#1}{\urlprefix }}%
\providecommand \urlprefix  [0]{URL }%
\providecommand \Eprint [0]{\href }%
\providecommand \doibase [0]{http://dx.doi.org/}%
\providecommand \selectlanguage [0]{\@gobble}%
\providecommand \bibinfo  [0]{\@secondoftwo}%
\providecommand \bibfield  [0]{\@secondoftwo}%
\providecommand \translation [1]{[#1]}%
\providecommand \BibitemOpen [0]{}%
\providecommand \bibitemStop [0]{}%
\providecommand \bibitemNoStop [0]{.\EOS\space}%
\providecommand \EOS [0]{\spacefactor3000\relax}%
\providecommand \BibitemShut  [1]{\csname bibitem#1\endcsname}%
\let\auto@bib@innerbib\@empty
%</preamble>
\bibitem [{\citenamefont {Tan}\ \emph {et~al.}(2016)\citenamefont {Tan} \emph
  {et~al.}}]{Tan:2016zwf}%
  \BibitemOpen
  \bibfield  {author} {\bibinfo {author} {\bibfnamefont {A.}~\bibnamefont
  {Tan}} \emph {et~al.} (\bibinfo {collaboration} {PandaX-II}),\ }\href
  {\doibase 10.1103/PhysRevLett.117.121303} {\bibfield  {journal} {\bibinfo
  {journal} {Phys. Rev. Lett.}\ }\textbf {\bibinfo {volume} {117}},\ \bibinfo
  {pages} {121303} (\bibinfo {year} {2016})},\ \Eprint
  {http://arxiv.org/abs/1607.07400} {arXiv:1607.07400 [hep-ex]} \BibitemShut
  {NoStop}%
%%CITATION = ARXIV:1607.07400;%%
\bibitem [{\citenamefont {Akerib}\ \emph {et~al.}(2016)\citenamefont {Akerib}
  \emph {et~al.}}]{Akerib:2016vxi}%
  \BibitemOpen
  \bibfield  {author} {\bibinfo {author} {\bibfnamefont {D.~S.}\ \bibnamefont
  {Akerib}} \emph {et~al.},\ }\href@noop {} {\  (\bibinfo {year} {2016})},\
  \Eprint {http://arxiv.org/abs/1608.07648} {arXiv:1608.07648 [astro-ph.CO]}
  \BibitemShut {NoStop}%
%%CITATION = ARXIV:1608.07648;%%
\bibitem [{\citenamefont {Aprile}\ \emph {et~al.}(2016)\citenamefont {Aprile}
  \emph {et~al.}}]{Aprile:2015uzo}%
  \BibitemOpen
  \bibfield  {author} {\bibinfo {author} {\bibfnamefont {E.}~\bibnamefont
  {Aprile}} \emph {et~al.} (\bibinfo {collaboration} {XENON}),\ }\href
  {\doibase 10.1088/1475-7516/2016/04/027} {\bibfield  {journal} {\bibinfo
  {journal} {JCAP}\ }\textbf {\bibinfo {volume} {1604}},\ \bibinfo {pages}
  {027} (\bibinfo {year} {2016})},\ \Eprint {http://arxiv.org/abs/1512.07501}
  {arXiv:1512.07501 [physics.ins-det]} \BibitemShut {NoStop}%
%%CITATION = ARXIV:1512.07501;%%
\bibitem [{\citenamefont {Akerib}\ \emph {et~al.}(2015)\citenamefont {Akerib}
  \emph {et~al.}}]{Akerib:2015cja}%
  \BibitemOpen
  \bibfield  {author} {\bibinfo {author} {\bibfnamefont {D.~S.}\ \bibnamefont
  {Akerib}} \emph {et~al.} (\bibinfo {collaboration} {LZ}),\ }\href@noop {} {\
  (\bibinfo {year} {2015})},\ \Eprint {http://arxiv.org/abs/1509.02910}
  {arXiv:1509.02910 [physics.ins-det]} \BibitemShut {NoStop}%
%%CITATION = ARXIV:1509.02910;%%
\bibitem [{\citenamefont {Aalbers}\ \emph {et~al.}(2016)\citenamefont {Aalbers}
  \emph {et~al.}}]{Aalbers:2016jon}%
  \BibitemOpen
  \bibfield  {author} {\bibinfo {author} {\bibfnamefont {J.}~\bibnamefont
  {Aalbers}} \emph {et~al.} (\bibinfo {collaboration} {DARWIN}),\ }\href@noop
  {} {\  (\bibinfo {year} {2016})},\ \Eprint {http://arxiv.org/abs/1606.07001}
  {arXiv:1606.07001 [astro-ph.IM]} \BibitemShut {NoStop}%
%%CITATION = ARXIV:1606.07001;%%
\bibitem [{\citenamefont {Billard}\ \emph {et~al.}(2014)\citenamefont
  {Billard}, \citenamefont {Strigari},\ and\ \citenamefont
  {Figueroa-Feliciano}}]{Billard:2013qya}%
  \BibitemOpen
  \bibfield  {author} {\bibinfo {author} {\bibfnamefont {J.}~\bibnamefont
  {Billard}}, \bibinfo {author} {\bibfnamefont {L.}~\bibnamefont {Strigari}}, \
  and\ \bibinfo {author} {\bibfnamefont {E.}~\bibnamefont
  {Figueroa-Feliciano}},\ }\href {\doibase 10.1103/PhysRevD.89.023524}
  {\bibfield  {journal} {\bibinfo  {journal} {Phys. Rev. D}\ }\textbf {\bibinfo
  {volume} {89}},\ \bibinfo {pages} {023524} (\bibinfo {year} {2014})},\
  \Eprint {http://arxiv.org/abs/1307.5458} {arXiv:1307.5458 [hep-ph]}
  \BibitemShut {NoStop}%
%%CITATION = ARXIV:1307.5458;%%
\bibitem [{\citenamefont {Baudis}\ \emph {et~al.}(2014)\citenamefont {Baudis},
  \citenamefont {Ferella}, \citenamefont {Kish}, \citenamefont {Manalaysay},
  \citenamefont {Marrodan~Undagoitia} \emph {et~al.}}]{Baudis:2013qla}%
  \BibitemOpen
  \bibfield  {author} {\bibinfo {author} {\bibfnamefont {L.}~\bibnamefont
  {Baudis}}, \bibinfo {author} {\bibfnamefont {A.}~\bibnamefont {Ferella}},
  \bibinfo {author} {\bibfnamefont {A.}~\bibnamefont {Kish}}, \bibinfo {author}
  {\bibfnamefont {A.}~\bibnamefont {Manalaysay}}, \bibinfo {author}
  {\bibfnamefont {T.}~\bibnamefont {Marrodan~Undagoitia}},  \emph {et~al.},\
  }\href {\doibase 10.1088/1475-7516/2014/01/044} {\bibfield  {journal}
  {\bibinfo  {journal} {JCAP}\ }\textbf {\bibinfo {volume} {1401}},\ \bibinfo
  {pages} {044} (\bibinfo {year} {2014})},\ \Eprint
  {http://arxiv.org/abs/1309.7024} {arXiv:1309.7024 [physics.ins-det]}
  \BibitemShut {NoStop}%
%%CITATION = ARXIV:1309.7024;%%
\bibitem [{\citenamefont {Bellini}\ \emph {et~al.}(2014)\citenamefont {Bellini}
  \emph {et~al.}}]{Bellini:2014uqa}%
  \BibitemOpen
  \bibfield  {author} {\bibinfo {author} {\bibfnamefont {G.}~\bibnamefont
  {Bellini}} \emph {et~al.} (\bibinfo {collaboration} {BOREXINO
  Collaboration}),\ }\href {\doibase 10.1038/nature13702} {\bibfield  {journal}
  {\bibinfo  {journal} {Nature}\ }\textbf {\bibinfo {volume} {512}},\ \bibinfo
  {pages} {383} (\bibinfo {year} {2014})}\BibitemShut {NoStop}%
%%CITATION = NATUA,512,383;%%
\bibitem [{\citenamefont {Bellini}\ \emph {et~al.}(2011)\citenamefont {Bellini}
  \emph {et~al.}}]{Bellini:2011rx}%
  \BibitemOpen
  \bibfield  {author} {\bibinfo {author} {\bibfnamefont {G.}~\bibnamefont
  {Bellini}} \emph {et~al.},\ }\href {\doibase 10.1103/PhysRevLett.107.141302}
  {\bibfield  {journal} {\bibinfo  {journal} {Phys. Rev. Lett.}\ }\textbf
  {\bibinfo {volume} {107}},\ \bibinfo {pages} {141302} (\bibinfo {year}
  {2011})},\ \Eprint {http://arxiv.org/abs/1104.1816} {arXiv:1104.1816
  [hep-ex]} \BibitemShut {NoStop}%
%%CITATION = ARXIV:1104.1816;%%
\bibitem [{\citenamefont {Chen}\ \emph
  {et~al.}(2014{\natexlab{a}})\citenamefont {Chen}, \citenamefont {Chi},
  \citenamefont {Huang}, \citenamefont {Liu}, \citenamefont {Shiao} \emph
  {et~al.}}]{Chen:2013lba}%
  \BibitemOpen
  \bibfield  {author} {\bibinfo {author} {\bibfnamefont {J.-W.}\ \bibnamefont
  {Chen}}, \bibinfo {author} {\bibfnamefont {H.-C.}\ \bibnamefont {Chi}},
  \bibinfo {author} {\bibfnamefont {K.-N.}\ \bibnamefont {Huang}}, \bibinfo
  {author} {\bibfnamefont {C.-P.}\ \bibnamefont {Liu}}, \bibinfo {author}
  {\bibfnamefont {H.-T.}\ \bibnamefont {Shiao}},  \emph {et~al.},\ }\href
  {\doibase 10.1016/j.physletb.2014.02.036} {\bibfield  {journal} {\bibinfo
  {journal} {Phys. Lett. B}\ }\textbf {\bibinfo {volume} {731}},\ \bibinfo
  {pages} {159} (\bibinfo {year} {2014}{\natexlab{a}})},\ \Eprint
  {http://arxiv.org/abs/1311.5294} {arXiv:1311.5294 [hep-ph]} \BibitemShut
  {NoStop}%
%%CITATION = ARXIV:1311.5294;%%
\bibitem [{\citenamefont {Chen}\ \emph
  {et~al.}(2014{\natexlab{b}})\citenamefont {Chen}, \citenamefont {Chi},
  \citenamefont {Li}, \citenamefont {Liu}, \citenamefont {Singh}, \citenamefont
  {Wong}, \citenamefont {Wu},\ and\ \citenamefont {Wu}}]{Chen:2014dsa}%
  \BibitemOpen
  \bibfield  {author} {\bibinfo {author} {\bibfnamefont {J.-W.}\ \bibnamefont
  {Chen}}, \bibinfo {author} {\bibfnamefont {H.-C.}\ \bibnamefont {Chi}},
  \bibinfo {author} {\bibfnamefont {H.-B.}\ \bibnamefont {Li}}, \bibinfo
  {author} {\bibfnamefont {C.~P.}\ \bibnamefont {Liu}}, \bibinfo {author}
  {\bibfnamefont {L.}~\bibnamefont {Singh}}, \bibinfo {author} {\bibfnamefont
  {H.~T.}\ \bibnamefont {Wong}}, \bibinfo {author} {\bibfnamefont {C.-L.}\
  \bibnamefont {Wu}}, \ and\ \bibinfo {author} {\bibfnamefont {C.-P.}\
  \bibnamefont {Wu}},\ }\href {\doibase 10.1103/PhysRevD.90.011301} {\bibfield
  {journal} {\bibinfo  {journal} {Phys. Rev. D}\ }\textbf {\bibinfo {volume}
  {90}},\ \bibinfo {pages} {011301} (\bibinfo {year} {2014}{\natexlab{b}})},\
  \Eprint {http://arxiv.org/abs/1405.7168} {arXiv:1405.7168 [hep-ph]}
  \BibitemShut {NoStop}%
%%CITATION = ARXIV:1405.7168;%%
\bibitem [{\citenamefont {Chen}\ \emph {et~al.}(2015)\citenamefont {Chen},
  \citenamefont {Chi}, \citenamefont {Huang}, \citenamefont {Li}, \citenamefont
  {Liu}, \citenamefont {Singh}, \citenamefont {Wong}, \citenamefont {Wu},\ and\
  \citenamefont {Wu}}]{Chen:2014ypv}%
  \BibitemOpen
  \bibfield  {author} {\bibinfo {author} {\bibfnamefont {J.-W.}\ \bibnamefont
  {Chen}}, \bibinfo {author} {\bibfnamefont {H.-C.}\ \bibnamefont {Chi}},
  \bibinfo {author} {\bibfnamefont {K.-N.}\ \bibnamefont {Huang}}, \bibinfo
  {author} {\bibfnamefont {H.-B.}\ \bibnamefont {Li}}, \bibinfo {author}
  {\bibfnamefont {C.-P.}\ \bibnamefont {Liu}}, \bibinfo {author} {\bibfnamefont
  {L.}~\bibnamefont {Singh}}, \bibinfo {author} {\bibfnamefont {H.~T.}\
  \bibnamefont {Wong}}, \bibinfo {author} {\bibfnamefont {C.-L.}\ \bibnamefont
  {Wu}}, \ and\ \bibinfo {author} {\bibfnamefont {C.-P.}\ \bibnamefont {Wu}},\
  }\href {\doibase 10.1103/PhysRevD.91.013005} {\bibfield  {journal} {\bibinfo
  {journal} {Phys. Rev. D}\ }\textbf {\bibinfo {volume} {91}},\ \bibinfo
  {pages} {013005} (\bibinfo {year} {2015})},\ \Eprint
  {http://arxiv.org/abs/1411.0574} {arXiv:1411.0574 [hep-ph]} \BibitemShut
  {NoStop}%
%%CITATION = ARXIV:1411.0574;%%
\bibitem [{\citenamefont {Johnson}\ and\ \citenamefont
  {Lin}(1979)}]{Johnson:1979wr}%
  \BibitemOpen
  \bibfield  {author} {\bibinfo {author} {\bibfnamefont {W.~R.}\ \bibnamefont
  {Johnson}}\ and\ \bibinfo {author} {\bibfnamefont {C.~D.}\ \bibnamefont
  {Lin}},\ }\href {\doibase 10.1103/PhysRevA.20.964} {\bibfield  {journal}
  {\bibinfo  {journal} {Phys. Rev. A}\ }\textbf {\bibinfo {volume} {20}},\
  \bibinfo {pages} {964} (\bibinfo {year} {1979})}\BibitemShut {NoStop}%
\bibitem [{\citenamefont {{Johnson}}\ and\ \citenamefont
  {{Cheng}}(1979)}]{Johnson:1979ch}%
  \BibitemOpen
  \bibfield  {author} {\bibinfo {author} {\bibfnamefont {W.~R.}\ \bibnamefont
  {{Johnson}}}\ and\ \bibinfo {author} {\bibfnamefont {K.~T.}\ \bibnamefont
  {{Cheng}}},\ }\href {\doibase 10.1103/PhysRevA.20.978} {\bibfield  {journal}
  {\bibinfo  {journal} {Phys. Rev. A}\ }\textbf {\bibinfo {volume} {20}},\
  \bibinfo {pages} {978} (\bibinfo {year} {1979})}\BibitemShut {NoStop}%
\bibitem [{\citenamefont {{Huang}}\ \emph {et~al.}(1981)\citenamefont
  {{Huang}}, \citenamefont {{Johnson}},\ and\ \citenamefont
  {{Cheng}}}]{Huang:1981jc}%
  \BibitemOpen
  \bibfield  {author} {\bibinfo {author} {\bibfnamefont {K.-N.}\ \bibnamefont
  {{Huang}}}, \bibinfo {author} {\bibfnamefont {W.~R.}\ \bibnamefont
  {{Johnson}}}, \ and\ \bibinfo {author} {\bibfnamefont {K.~T.}\ \bibnamefont
  {{Cheng}}},\ }\href {\doibase 10.1016/0092-640X(81)90010-3} {\bibfield
  {journal} {\bibinfo  {journal} {At. Data Nucl. Data Tables}\ }\textbf
  {\bibinfo {volume} {26}},\ \bibinfo {pages} {33} (\bibinfo {year}
  {1981})}\BibitemShut {NoStop}%
\bibitem [{\citenamefont {{Huang}}\ and\ \citenamefont
  {{Johnson}}(1982)}]{Huang:1981wj}%
  \BibitemOpen
  \bibfield  {author} {\bibinfo {author} {\bibfnamefont {K.-N.}\ \bibnamefont
  {{Huang}}}\ and\ \bibinfo {author} {\bibfnamefont {W.~R.}\ \bibnamefont
  {{Johnson}}},\ }\href {\doibase 10.1103/PhysRevA.25.634} {\bibfield
  {journal} {\bibinfo  {journal} {Phys. Rev. A}\ }\textbf {\bibinfo {volume}
  {25}},\ \bibinfo {pages} {634} (\bibinfo {year} {1982})}\BibitemShut
  {NoStop}%
\bibitem [{\citenamefont {{Henke}}\ \emph {et~al.}(1993)\citenamefont
  {{Henke}}, \citenamefont {{Gullikson}},\ and\ \citenamefont
  {{Davis}}}]{Henke:1993gd}%
  \BibitemOpen
  \bibfield  {author} {\bibinfo {author} {\bibfnamefont {B.~L.}\ \bibnamefont
  {{Henke}}}, \bibinfo {author} {\bibfnamefont {E.~M.}\ \bibnamefont
  {{Gullikson}}}, \ and\ \bibinfo {author} {\bibfnamefont {J.~C.}\ \bibnamefont
  {{Davis}}},\ }\href {\doibase 10.1006/adnd.1993.1013} {\bibfield  {journal}
  {\bibinfo  {journal} {Atom. Data Nucl. Data Tables}\ }\textbf {\bibinfo
  {volume} {54}},\ \bibinfo {pages} {181} (\bibinfo {year} {1993})}\BibitemShut
  {NoStop}%
\bibitem [{\citenamefont {Samson}\ and\ \citenamefont
  {Stolte}(2002)}]{Samson:2002xe}%
  \BibitemOpen
  \bibfield  {author} {\bibinfo {author} {\bibfnamefont {J.}~\bibnamefont
  {Samson}}\ and\ \bibinfo {author} {\bibfnamefont {W.}~\bibnamefont
  {Stolte}},\ }\href {\doibase http://dx.doi.org/10.1016/S0368-2048(02)00026-9}
  {\bibfield  {journal} {\bibinfo  {journal} {J. Electron Spectrosc. Relat.
  Phenom.}\ }\textbf {\bibinfo {volume} {123}},\ \bibinfo {pages} {265 }
  (\bibinfo {year} {2002})}\BibitemShut {NoStop}%
\bibitem [{\citenamefont {Suzuki}\ and\ \citenamefont
  {Saito}(2003)}]{Suzuki:2003xe}%
  \BibitemOpen
  \bibfield  {author} {\bibinfo {author} {\bibfnamefont {I.~H.}\ \bibnamefont
  {Suzuki}}\ and\ \bibinfo {author} {\bibfnamefont {N.}~\bibnamefont {Saito}},\
  }\href {\doibase http://dx.doi.org/10.1016/S0368-2048(03)00012-4} {\bibfield
  {journal} {\bibinfo  {journal} {J. Electron Spectrosc. Relat. Phenom.}\
  }\textbf {\bibinfo {volume} {129}},\ \bibinfo {pages} {71 } (\bibinfo {year}
  {2003})}\BibitemShut {NoStop}%
\bibitem [{\citenamefont {Zheng}\ \emph {et~al.}(2006)\citenamefont {Zheng},
  \citenamefont {Cui}, \citenamefont {Zhao}, \citenamefont {Zhao},\ and\
  \citenamefont {Chen}}]{Zheng:2006xe}%
  \BibitemOpen
  \bibfield  {author} {\bibinfo {author} {\bibfnamefont {L.}~\bibnamefont
  {Zheng}}, \bibinfo {author} {\bibfnamefont {M.}~\bibnamefont {Cui}}, \bibinfo
  {author} {\bibfnamefont {Y.}~\bibnamefont {Zhao}}, \bibinfo {author}
  {\bibfnamefont {J.}~\bibnamefont {Zhao}}, \ and\ \bibinfo {author}
  {\bibfnamefont {K.}~\bibnamefont {Chen}},\ }\href {\doibase
  http://dx.doi.org/10.1016/j.elspec.2006.04.006} {\bibfield  {journal}
  {\bibinfo  {journal} {J. Electron Spectrosc. Relat. Phenom.}\ }\textbf
  {\bibinfo {volume} {152}},\ \bibinfo {pages} {143 } (\bibinfo {year}
  {2006})}\BibitemShut {NoStop}%
\bibitem [{\citenamefont {{Band}}\ \emph {et~al.}(1979)\citenamefont {{Band}},
  \citenamefont {{Kharitonov}},\ and\ \citenamefont
  {{Trzhask-Ovskaya}}}]{Band:1979xe}%
  \BibitemOpen
  \bibfield  {author} {\bibinfo {author} {\bibfnamefont {M.}~\bibnamefont
  {{Band}}}, \bibinfo {author} {\bibfnamefont {Y.~I.}\ \bibnamefont
  {{Kharitonov}}}, \ and\ \bibinfo {author} {\bibfnamefont {M.~B.}\
  \bibnamefont {{Trzhask-Ovskaya}}},\ }\href {\doibase
  10.1016/0092-640X(79)90027-5} {\bibfield  {journal} {\bibinfo  {journal} {At.
  Data Nucl. Data Tables}\ }\textbf {\bibinfo {volume} {23}},\ \bibinfo {pages}
  {443} (\bibinfo {year} {1979})}\BibitemShut {NoStop}%
\bibitem [{\citenamefont {{Yeh}}\ and\ \citenamefont
  {{Lindau}}(1985)}]{Yeh:1985xe}%
  \BibitemOpen
  \bibfield  {author} {\bibinfo {author} {\bibfnamefont {J.~J.}\ \bibnamefont
  {{Yeh}}}\ and\ \bibinfo {author} {\bibfnamefont {I.}~\bibnamefont
  {{Lindau}}},\ }\href {\doibase 10.1016/0092-640X(85)90016-6} {\bibfield
  {journal} {\bibinfo  {journal} {At. Data Nucl. Data Tables}\ }\textbf
  {\bibinfo {volume} {32}},\ \bibinfo {pages} {1} (\bibinfo {year}
  {1985})}\BibitemShut {NoStop}%
\bibitem [{\citenamefont {Serenelli}\ \emph {et~al.}(2011)\citenamefont
  {Serenelli}, \citenamefont {Haxton},\ and\ \citenamefont
  {Pena-Garay}}]{Serenelli:2011py}%
  \BibitemOpen
  \bibfield  {author} {\bibinfo {author} {\bibfnamefont {A.~M.}\ \bibnamefont
  {Serenelli}}, \bibinfo {author} {\bibfnamefont {W.~C.}\ \bibnamefont
  {Haxton}}, \ and\ \bibinfo {author} {\bibfnamefont {C.}~\bibnamefont
  {Pena-Garay}},\ }\href {\doibase 10.1088/0004-637X/743/1/24} {\bibfield
  {journal} {\bibinfo  {journal} {Astrophys. J.}\ }\textbf {\bibinfo {volume}
  {743}},\ \bibinfo {pages} {24} (\bibinfo {year} {2011})},\ \Eprint
  {http://arxiv.org/abs/1104.1639} {arXiv:1104.1639 [astro-ph.SR]} \BibitemShut
  {NoStop}%
%%CITATION = ARXIV:1104.1639;%%
\bibitem [{\citenamefont {Adelberger}\ \emph {et~al.}(2011)\citenamefont
  {Adelberger} \emph {et~al.}}]{Adelberger:2010qa}%
  \BibitemOpen
  \bibfield  {author} {\bibinfo {author} {\bibfnamefont {E.~G.}\ \bibnamefont
  {Adelberger}} \emph {et~al.},\ }\href {\doibase 10.1103/RevModPhys.83.195}
  {\bibfield  {journal} {\bibinfo  {journal} {Rev. Mod. Phys.}\ }\textbf
  {\bibinfo {volume} {83}},\ \bibinfo {pages} {195} (\bibinfo {year} {2011})},\
  \Eprint {http://arxiv.org/abs/1004.2318} {arXiv:1004.2318 [nucl-ex]}
  \BibitemShut {NoStop}%
%%CITATION = ARXIV:1004.2318;%%
\bibitem [{\citenamefont {Bahcall}\ \emph {et~al.}(2005)\citenamefont
  {Bahcall}, \citenamefont {Serenelli},\ and\ \citenamefont
  {Basu}}]{Bahcall:2004pz}%
  \BibitemOpen
  \bibfield  {author} {\bibinfo {author} {\bibfnamefont {J.~N.}\ \bibnamefont
  {Bahcall}}, \bibinfo {author} {\bibfnamefont {A.~M.}\ \bibnamefont
  {Serenelli}}, \ and\ \bibinfo {author} {\bibfnamefont {S.}~\bibnamefont
  {Basu}},\ }\href {\doibase 10.1086/428929} {\bibfield  {journal} {\bibinfo
  {journal} {Astrophys. J.}\ }\textbf {\bibinfo {volume} {621}},\ \bibinfo
  {pages} {L85} (\bibinfo {year} {2005})},\ \Eprint
  {http://arxiv.org/abs/astro-ph/0412440} {arXiv:astro-ph/0412440 [astro-ph]}
  \BibitemShut {NoStop}%
%%CITATION = ASTRO-PH/0412440;%%
\bibitem [{\citenamefont {Bahcall}(1997)}]{Bahcall:1997eg}%
  \BibitemOpen
  \bibfield  {author} {\bibinfo {author} {\bibfnamefont {J.~N.}\ \bibnamefont
  {Bahcall}},\ }\href {\doibase 10.1103/PhysRevC.56.3391} {\bibfield  {journal}
  {\bibinfo  {journal} {Phys. Rev. C}\ }\textbf {\bibinfo {volume} {56}},\
  \bibinfo {pages} {3391} (\bibinfo {year} {1997})},\ \Eprint
  {http://arxiv.org/abs/hep-ph/9710491} {arXiv:hep-ph/9710491 [hep-ph]}
  \BibitemShut {NoStop}%
%%CITATION = HEP-PH/9710491;%%
\bibitem [{\citenamefont {Bahcall}\ and\ \citenamefont
  {Pena-Garay}(2004)}]{Bahcall:2004mz}%
  \BibitemOpen
  \bibfield  {author} {\bibinfo {author} {\bibfnamefont {J.~N.}\ \bibnamefont
  {Bahcall}}\ and\ \bibinfo {author} {\bibfnamefont {C.}~\bibnamefont
  {Pena-Garay}},\ }\href {\doibase 10.1088/1367-2630/6/1/063} {\bibfield
  {journal} {\bibinfo  {journal} {New J. Phys.}\ }\textbf {\bibinfo {volume}
  {6}},\ \bibinfo {pages} {63} (\bibinfo {year} {2004})},\ \Eprint
  {http://arxiv.org/abs/hep-ph/0404061} {arXiv:hep-ph/0404061 [hep-ph]}
  \BibitemShut {NoStop}%
%%CITATION = HEP-PH/0404061;%%
\bibitem [{\citenamefont {Olive}\ \emph {et~al.}(2014)\citenamefont {Olive}
  \emph {et~al.}}]{Agashe:2014kda}%
  \BibitemOpen
  \bibfield  {author} {\bibinfo {author} {\bibfnamefont {K.~A.}\ \bibnamefont
  {Olive}} \emph {et~al.} (\bibinfo {collaboration} {Particle Data Group}),\
  }\href {\doibase 10.1088/1674-1137/38/9/090001} {\bibfield  {journal}
  {\bibinfo  {journal} {Chin. Phys. C}\ }\textbf {\bibinfo {volume} {38}},\
  \bibinfo {pages} {090001} (\bibinfo {year} {2014})}\BibitemShut {NoStop}%
%%CITATION = CHPHD,C38,090001;%%
\bibitem [{\citenamefont {Akerib}\ \emph {et~al.}(2014)\citenamefont {Akerib}
  \emph {et~al.}}]{Akerib:2013tjd}%
  \BibitemOpen
  \bibfield  {author} {\bibinfo {author} {\bibfnamefont {D.~S.}\ \bibnamefont
  {Akerib}} \emph {et~al.} (\bibinfo {collaboration} {LUX}),\ }\href {\doibase
  10.1103/PhysRevLett.112.091303} {\bibfield  {journal} {\bibinfo  {journal}
  {Phys. Rev. Lett.}\ }\textbf {\bibinfo {volume} {112}},\ \bibinfo {pages}
  {091303} (\bibinfo {year} {2014})},\ \Eprint {http://arxiv.org/abs/1310.8214}
  {arXiv:1310.8214 [astro-ph.CO]} \BibitemShut {NoStop}%
%%CITATION = ARXIV:1310.8214;%%
\bibitem [{\citenamefont {Akimov}\ \emph {et~al.}(2012)\citenamefont {Akimov}
  \emph {et~al.}}]{Akimov:2011tj}%
  \BibitemOpen
  \bibfield  {author} {\bibinfo {author} {\bibfnamefont {D.~{\relax Yu}.}\
  \bibnamefont {Akimov}} \emph {et~al.},\ }\href {\doibase
  10.1016/j.physletb.2012.01.064} {\bibfield  {journal} {\bibinfo  {journal}
  {Phys. Lett.}\ }\textbf {\bibinfo {volume} {B709}},\ \bibinfo {pages} {14}
  (\bibinfo {year} {2012})},\ \Eprint {http://arxiv.org/abs/1110.4769}
  {arXiv:1110.4769 [astro-ph.CO]} \BibitemShut {NoStop}%
%%CITATION = ARXIV:1110.4769;%%
\bibitem [{\citenamefont {Schumann}\ \emph {et~al.}(2015)\citenamefont
  {Schumann}, \citenamefont {Baudis}, \citenamefont {Bütikofer}, \citenamefont
  {Kish},\ and\ \citenamefont {Selvi}}]{Schumann:2015cpa}%
  \BibitemOpen
  \bibfield  {author} {\bibinfo {author} {\bibfnamefont {M.}~\bibnamefont
  {Schumann}}, \bibinfo {author} {\bibfnamefont {L.}~\bibnamefont {Baudis}},
  \bibinfo {author} {\bibfnamefont {L.}~\bibnamefont {Bütikofer}}, \bibinfo
  {author} {\bibfnamefont {A.}~\bibnamefont {Kish}}, \ and\ \bibinfo {author}
  {\bibfnamefont {M.}~\bibnamefont {Selvi}},\ }\href {\doibase
  10.1088/1475-7516/2015/10/016} {\bibfield  {journal} {\bibinfo  {journal}
  {JCAP}\ }\textbf {\bibinfo {volume} {1510}},\ \bibinfo {pages} {016}
  (\bibinfo {year} {2015})},\ \Eprint {http://arxiv.org/abs/1506.08309}
  {arXiv:1506.08309 [physics.ins-det]} \BibitemShut {NoStop}%
%%CITATION = ARXIV:1506.08309;%%
\end{thebibliography}%

\end{document}